\documentclass[aps,pra,superscriptaddress]{revtex4}
\usepackage{amsmath,epsfig}
\usepackage{graphicx}

\def\be{\begin{equation}}
\def\ee{\end{equation}}
\def\bea{\begin{eqnarray}}
\def\eea{\end{eqnarray}}
\newcommand{\ket}[1]{\mbox{$|#1\rangle$}}
\newcommand{\bra}[1]{\mbox{$\langle#1|$}}
\newcommand{\avg}[1]{\mbox{$\langle#1\rangle$}}

\newcommand{\opdagger}[2]{\mbox{$\hat{#1}_{#2}^{\dagger}$}}

\def\bfr{{\bf r}}

\def\omegapl{\omega_{\footnotesize\textrm{pl}}}
\def\Omegamin{\Omega_{\footnotesize\textrm{min}}}

\begin{document}

\title{Trapping atoms using nanoscale quantum vacuum forces}
\date{\today}

\author{D.E. Chang}
\affiliation{ICFO - Institut de Ciencies Fotoniques, Mediterranean
Technology Park, 08860 Castelldefels (Barcelona), Spain}
\email{darrick.chang@icfo.es}

\author{K. Sinha}
\affiliation{Joint Quantum Institute, College Park, MD 20742, USA}

\author{J.M. Taylor}
\affiliation{Joint Quantum Institute, College Park, MD 20742, USA}
\affiliation{National Institute of Standards and Technology, 100 Bureau Dr MS 8410, Gaithersburg, MD 20899, USA}

\author{H.J. Kimble}
\affiliation{IQIM, California Institute of Technology, Pasadena,
CA 91125, USA} \affiliation{Norman Bridge Laboratory of Physics
12-33, California Institute of Technology, Pasadena, CA 91125,
USA}

\def\bfk{{\bf k}}
\def\boldrho{\boldsymbol{\rho}}
\def\kperp{k_{\perp}}
\def\kp{k_{\parallel}}

\begin{abstract}
Quantum vacuum forces dictate the interaction between individual atoms and dielectric surfaces at nanoscale distances. For example, their large strengths typically overwhelm externally applied forces, which makes it challenging to controllably interface cold atoms with nearby nanophotonic systems. Here, we show that it is possible to tailor the vacuum forces themselves to provide strong trapping potentials. The trapping scheme takes advantage of the attractive ground state potential and adiabatic dressing with an excited state whose potential is engineered to be resonantly enhanced and repulsive. This procedure yields a strong metastable trap, with the fraction of excited state population scaling inversely with the quality factor of the resonance of the dielectric structure. We analyze realistic limitations to the trap lifetime and discuss possible applications that might emerge from the large trap depths and nanoscale confinement.
\end{abstract}
\maketitle


\section{Introduction}

One of the spectacular predictions of quantum electrodynamics is the emergence of forces that arise purely from quantum fluctuations of the electromagnetic vacuum~\cite{lamoreaux_casimir_2005}. Also known as London-van der Waals~\cite{london_general_1937} or Casimir forces~\cite{casimir_influence_1948} in different regimes, these forces are often dominant at short distances and can give rise to undesirable effects in nanoscale systems, such as nanomechanical stiction~\cite{buks_stiction_2001}. These forces have also attracted increasing attention in the fields of atomic physics and quantum optics. In particular, significant efforts have been made in recent years to interface cold atoms with the evanescent fields of dielectric micro- and nano-photonic systems~\cite{renn_laser-guided_1995,nayak_optical_2007,bajcsy_efficient_2009,vetsch_optical_2010,alton_strong_2010,stehle_plasmonically_2011,goban_demonstration_2012,thompson_coupling_2013}. These systems are expected to facilitate strong, tunable interactions between individual atoms and photons for applications such as quantum information processing~\cite{kimble_quantum_2008} and the investigation of quantum many-body physics~\cite{greentree_quantum_2006,hartmann_strongly_2006,chang_crystallization_2008}. In practice, efficient atomic coupling to the evanescent fields of the nanophotonic systems requires that atoms be trapped within sub-wavelength distances of these structures.

At these scales, quantum vacuum forces can overwhelm the forces associated with conventional optical dipole traps, typically resulting in a loss of trap stability at distances $d\lesssim 100$ nm from dielectric surfaces. Given the ability to engineer the properties of nanophotonic structures, an interesting question arises as to whether such systems could be used to significantly modify vacuum forces, perhaps changing their sign from being attractive to repulsive, or even creating local potential minima. The strength of nanoscale vacuum forces should lead to unprecedented energy and length scales for atomic traps, which would find use beyond nanophotonic interfaces, such as in quantum simulation protocols based upon ultracold atoms~\cite{lewenstein_ultracold_2012} and control of inter-atomic interactions~\cite{bolda_effective-scattering-length_2002}.

Here we propose a novel mechanism in which engineered vacuum forces can enable the formation of a nanoscale atomic trap. While a recent no-go theorem~\cite{rahi_constraints_2010} forbids a vacuum trap for atoms in their electronic ground states, tailored nanophotonic systems can yield strong repulsive potentials for atomic excited states. We show that a weak external optical field can give rise to an overall trapping potential for a dressed state. Remarkably, absent fundamental limits on the losses of the surrounding dielectric structure, the fraction of excited-state population in the dressed state can become infinitesimal, which greatly enhances the trapping lifetime and stability. We identify and carefully analyze the actual limiting mechanisms. While we present calculations on a simple model where analytical results can be obtained, we also discuss the generality of our protocol to realistic systems such as photonic crystal structures.

The no-go theorem for non-magnetic media states that a dielectric object in vacuum cannot be stably trapped with vacuum forces for any surrounding configuration of dielectric objects, provided that the system is in thermal equilibrium~\cite{rahi_constraints_2010}. In analogy with Earnshaw's theorem, which prohibits trapping of charged objects with static electric potentials, at best one can create a saddle-point potential~(see, \textit{e.g.}, Ref.~\cite{levin_casimir_2010} involving metal particles and Ref.~\cite{hung_trapped_2013} for a hybrid vacuum/optical trap for atoms). For dielectric objects, potential loopholes involve embedding the system in a high-index fluid~\cite{munday_measured_2009}, or using blackbody radiation pressure associated with strong temperature gradients~\cite{antezza_new_2005}. Applied to atoms, this theorem essentially forbids stable vacuum trapping near a dielectric structure when the atom is in its electronic ground state. No such constraint exists for atoms in their excited states~\cite{failache_resonant_1999}, although the robustness of any possible trap would be limited by the excited state lifetime. Here, we show that an atom only weakly dressed by its excited state can be trapped by properly tailoring the dispersion of the underlying structure.

\section{Atomic interactions with vacuum}

We begin by briefly reviewing how quantum vacuum fluctuations give rise to forces on the ground and excited states. Within an effective two-level approximation of an isotropic atom with ground and excited states $\ket{g},\ket{e}$~(see Appendix for description of how isotropy is incorporated into the model), the interaction between the electromagnetic field and an atom at position $\bfr$ is given by the dipole Hamiltonian $H=-{\bf d}\cdot{\bf E}({\bf r})$. Following a complete decomposition of the electric field into its normal modes $k$, one can write $H=-\sum_{k} g_k(\bfr)(\sigma_{eg}+\sigma_{ge})(\hat{a}_k+\opdagger{a}{k})$, where $\sigma_{ij}=\ket{i}\bra{j}$. The energy non-conserving terms~($\sigma_{eg}\opdagger{a}{k}$ and $\sigma_{ge}\hat{a}_k$) enable an atom in its ground state $\ket{g,0}$ to couple virtually to the excited state and create a photon, $\ket{e,1_k}$, which is subsequently re-absorbed. The corresponding frequency shift for the ground state within second-order perturbation theory is $\omega_{g}(\bfr)=-\sum_{k} g_k(\bfr)^2/(\omega_{0}+\omega_k)$, where $\omega_0$ is the unperturbed atomic transition frequency. This shift results in a mechanical ``vacuum'' potential when translational symmetry is broken due to dielectric surfaces. Using a quantization technique for electrodynamics in the presence of dispersive dielectric media, the shift can be expressed in terms of the scattered component of the dyadic electromagnetic Green's function evaluated at imaginary frequencies $\omega=iu$~\cite{buhmann_dispersion_2007}~(see Appendix),
\be \delta\omega_g({\bfr})=\frac{3c\Gamma_0}{\omega_0^2}\int_{0}^{\infty}du\frac{u^2}{\omega_0^2+u^2}\textrm{Tr}\,G_{\footnotesize\textrm{sc}}(\bfr,\bfr,iu), \ee
where $\Gamma_0$ is the free-space spontaneous emission rate of the atom.

A similar process can occur for an atom in its excited state $\ket{e,0}$, in which the atom virtually emits and re-absorbs an off-resonant photon~($\ket{g,1_k}$). However, the excited state is unique in that it can also emit a resonant photon. The total excited state shift is given by~\cite{buhmann_dispersion_2007}
\be \delta\omega_e(\bfr)=-\frac{\delta\omega_g(\bfr)}{3}-\frac{\Gamma_0 \pi c}{\omega_0}\textrm{Tr}\,\textrm{Re}\,G_{\footnotesize\textrm{sc}}(\bfr,\bfr,\omega_0).\label{eq:omegae} \ee
The first term on the right arises from off-resonant processes, while the second term can be interpreted as the interaction between the atom and its own resonantly emitted photon. Naturally, the dielectric environment can modify the spontaneous emission rate as well,
\be \Gamma(\bfr)=\Gamma_0+\frac{2\Gamma_0 \pi c}{\omega_0}\textrm{Tr}\,\textrm{Im}\,G_{\footnotesize\textrm{sc}}(\bfr,\bfr,\omega_0).\label{eq:Gamma} \ee
Note that the modified emission and resonant shifts are complementary in that they emerge from different quadratures of the Green's function. The versatility of micro- and nanophotonic systems has already been exploited to greatly modify emission rates of atoms and molecules in a variety of contexts~\cite{kneipp_surface-enhanced_2002,englund_controlling_2005,chang_quantum_2006,aoki_observation_2006}. We propose that these same systems can be used to tailor excited-state potentials in order to facilitate vacuum trapping. Of course, the excited-state shift can be affected by coupling to even higher electronic levels. However, since our goal is to engineer giant resonant shifts of $\ket{e}$, the off-resonant couplings to higher-lying states yield only minor corrections and consequently are ignored here.

\begin{figure}[t]
\begin{center}
\includegraphics[width=13cm]{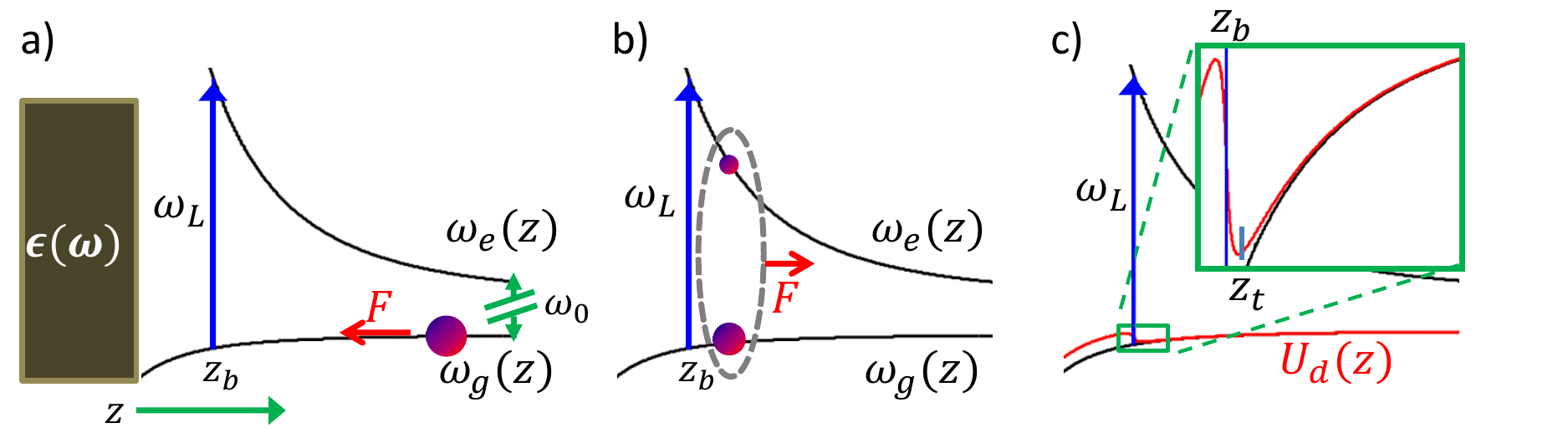}
\end{center}
\caption{\textbf{Schematic of vacuum trapping mechanism.} a) A semi-infinite dielectric with specially tailored frequency-dependent permittivity $\epsilon(\omega)$ enables large repulsion in the excited state $\omega_e(z)$, in contrast to the attractive vacuum potential experienced by an atom in its ground state. An overall trapping potential normal to the surface (along $z$) can be generated by driving the atom with an external laser of frequency $\omega_L$, which is greater than the natural resonance frequency $\omega_0$ of the atom. The laser comes into local resonance with the atom at $z_b$. An atom far from $z_b$ is unaffected by the largely detuned field and feels an attractive force $F$ toward the dielectric interface. b) The atom comes closer to local resonance with the laser as $z\rightarrow z_b$. This creates an atomic dressed state, whose excited state component yields a net repulsive force on the atom. c) A slowly moving atom experiences the adiabatic dressed state potential $U_d(z)$ shown in red, which characterizes the position-dependent mixing between ground and excited states. The minimum of the dressed potential is located at $z_t>z_b$.\label{fig:schematic}}
\end{figure}

In what follows, we present a simple model system that enables one-dimensional trapping in a plane parallel to a semi-infinite dielectric slab, as illustrated in Fig.~\ref{fig:schematic}. The use of a homogeneous dielectric~(as opposed to, \textit{e.g.}, a photonic crystal) greatly simplifies the calculation and enables one to understand the relevant trap properties analytically, although we later argue why the qualitative features should still hold in more complex geometries. In the dielectric slab model, dispersion engineering will be realized via the frequency-dependent electric permittivity $\epsilon(\omega)$. While this particular system achieves a one-dimensional trap along $z$, our electrodynamic calculations are performed in three dimensions.

A short-distance expansion of the Green's function, valid for sub-wavelength scales, shows that the excited-state emission rate $\Gamma(z)$ and resonant contribution $\delta\omega_e^{(r)}(z)$ to the shift are given by
\bea \frac{\Gamma(z)}{\Gamma_0} & \approx & \frac{1}{4(k_0 z)^3}\textrm{Im}\left(\frac{\epsilon_a-1}{\epsilon_a+1}\right),\label{eq:Gammageneral} \\ \frac{\delta\omega_{e}^{(r)}(z)}{\Gamma_0} & \approx & -\frac{1}{8(k_0 z)^3}\textrm{Re}\left(\frac{\epsilon_a-1}{\epsilon_a+1}\right),\label{eq:omegaegeneral} \eea
where the permittivity $\epsilon_a=\epsilon(\omega_0)$ is evaluated at the atomic resonance frequency, $z$ is the distance from the surface, and $k_0=\omega_0/c$ is the resonant free-space wavevector. A large repulsive potential is generated when $\epsilon_a\rightarrow-1^{+}$, which is analogous to the interaction between a classical dipole at position $z$ and its large induced image dipole in the dielectric. The modified spontaneous emission rate originates from the absorption or quenching of the atomic emission due to material losses.

We choose a Drude model, $\epsilon(\omega)=1-\frac{\omega_p^2}{\omega^2+i\omega\gamma}$, as a simple dielectric function. Note that this function satisfies Kramers-Kronig relations~(causality), as a choice of a non-causal function could result in an apparent violation of the no-go theorem~\cite{rahi_constraints_2010} as well. In the limit of vanishing material loss parameter $\gamma$, the system passes through $\epsilon=-1$ at the plasmon resonance frequency $\omegapl=\omega_p/\sqrt{2}$. The system is conveniently parameterized by the quality factor $Q\equiv\omegapl/\gamma$ and a dimensionless detuning $\Delta_p=(\omega-\omegapl)/\gamma$. Then, for large $Q$ and $\Delta_p\gg 1$, $\frac{\Gamma}{\Gamma_0}\approx\frac{1}{16(k_0 z)^3}\frac{Q}{\Delta_p^2}$ and $\frac{\delta\omega_e^{(r)}}{\Gamma_0}\approx\frac{1}{16(k_0 z)^3}\frac{Q}{\Delta_p}$. The dispersion and dissipation scale like $\Delta_p^{-1}$ and $\Delta_p^{-2}$, respectively. Thus, for high $Q$, it is possible to choose detunings where the atom still sees significant repulsive forces, but where spontaneous emission into the material is not significantly enhanced~(in addition to material-induced emission of Eq.~(\ref{eq:Gammageneral}), there still is emission into free-space at a rate $\sim\Gamma_0$). This scaling behavior is in close analogy with conventional optical trapping of atoms, where simultaneously using large trapping intensity and detuning maintains reasonable trap depths but suppresses unwanted photon scattering. The ground-state shift for atomic frequencies $\omega_{0}\sim\omegapl$ and in the near-field is given by $\delta\omega_g\approx-\frac{3\Gamma_0}{32(k_0 z)^3}$.

\section{A nanoscale vacuum force trap}

We now describe how a trap can form for an atom subject to these potentials, and in the presence of a weak driving laser of frequency $\omega_L$ and Rabi frequency $\Omega$~(see Fig.~\ref{fig:schematic}). For conceptual simplicity, we choose the Rabi frequency to be spatially uniform, such that all atomic forces are attributable to the vacuum potentials alone. Qualitatively, the strong position dependence of the ground and excited state frequencies causes the laser to come into resonance with the atom at a single, tunable point $z=z_b$. Under certain conditions~(described in detail later), an atom starting at position $z\gg z_b$ will be far detuned, such that it is essentially in the ground state and is attracted by the pure ground-state potential. However, moving closer to $z_b$ brings the atom closer to resonance. The dressing of the atom with a small fraction of excited state population causes a repulsive barrier to form near $z\sim z_b$, yielding a metastable trap. Significantly, our approach does not attempt to directly counteract the attractive ground-state potential with large external optical potentials, in sharp contrast with other trapping schemes near dielectric surfaces~\cite{vetsch_optical_2010,alton_strong_2010,murphy_electro-optical_2009,goban_demonstration_2012,chang_trapping_2009,gullans_nanoplasmonic_2012,thompson_coupling_2013}.

The order of our calculation of the trap properties is as follows. We first find the adiabatic potential experienced by a slowly-moving atom. We derive relevant properties around the location of the trap minimum $z_t$~(Fig.~\ref{fig:schematic}). In particular, we show that absent material constraints~(arbitrarily high $Q$), the dressed state can in principle have infinitesimal excited state population and scattering. We then quantize the motion to find the motional eigenstates, binding energy, and position uncertainty $\Delta z$. Finally, we analyze the mechanisms that limit the trap lifetime.

First treating the motion classically, the force experienced by the atom is given by $\frac{dp}{dt}=-\hbar(\frac{d\omega_e}{dz}\sigma_{ee}(t)+\frac{d\omega_g}{dz}\sigma_{gg}(t))$, while the internal atomic dynamics satisfies the usual Bloch equations, \textit{e.g.},
\be \frac{d\sigma_{ge}}{dt}=(i\delta_a(z)-\Gamma(z)/2)\sigma_{ge}+i\frac{\Omega}{2}(\sigma_{ee}-\sigma_{gg})\label{eq:Bloch}. \ee
Here $\delta_a(z)=\omega_L-(\omega_e(z)-\omega_g(z))$ is the detuning between the laser and the local atomic resonance frequency. When the atom moves slowly on the time scales of the internal dynamics, the atomic coherence can be adiabatically eliminated, $\frac{d\sigma_{ge}}{dt}\approx 0$, such that the atomic populations are functions of position alone, $\sigma_{ee}=\frac{\Omega^2}{\Gamma(z)^2+4\delta_a(z)^2+2\Omega^2}$. We are primarily interested in the regime where the atom is weakly driven, $\sigma_{ee}\ll 1$. A more careful calculation reveals that the Green's function $G_{\footnotesize\textrm{sc}}(\bfr,\bfr,\omega)$ for the excited state resonant shift and emission rate in Eqs.~(\ref{eq:omegae}) and~(\ref{eq:Gamma}) must in fact be evaluated at the laser frequency in this regime, which reflects that the atom primarily acts as a Rayleigh scatterer~(see Appendix). Consequently, $\epsilon_a$ in Eqs.~(\ref{eq:Gammageneral}) and~(\ref{eq:omegaegeneral}) is replaced with $\epsilon(\omega_L)$. The adiabatic potential seen by a slowly-moving atom is $U_d(z)=\int_{\infty}^{z}d\tilde{z}\frac{dp(\tilde{z})}{dt}$.

Three independent frequencies characterize the system~($\omega_L,\omega_p$, and $\omega_{0}$), and two relative frequencies must be specified. One is determined by selecting the trap barrier position, determined by the relation $\delta_a(z_b)\equiv 0$. A second is the dimensionless detuning between the plasmon resonance and laser, $\Delta_p=(\omega_L-\omegapl)/\gamma$, which we treat as a free parameter that determines the relative strengths of the excited state shift and dissipation. In the near-field, the excited state repulsion exceeds the ground state attraction by a factor $2Q/3\Delta_p$. The population at $z_b$ must then exceed the inverse of this quantity, $\sigma_{ee}>3\Delta_p/2Q$, in order to provide a classical barrier in the adiabatic potential, which translates to a minimum Rabi frequency $\frac{\Omegamin}{\Gamma_0}\propto\frac{Q^{1/2}}{\Delta_p^{3/2}(k_0 z_b)^3}$. Since $\Delta_p\lesssim Q$, this scheme potentially enables atom trapping near surfaces with greatly reduced intensities as compared to conventional trapping. In the latter case, an optical dipole potential requires a minimum driving field of $\frac{\Omegamin}{\Gamma_0}\sim \frac{1}{(k_0 z_b)^3}$ to overcome the attractive ground state potential.

The excited-state population at the trap minimum is fixed by the ratio of ground to excited state forces, $\sigma_{ee}(z_t)=\frac{3\Delta_p}{2Q}$, and is surprisingly independent of $\Omega$~(provided that $\Omega>\Omegamin$). An increasing Rabi frequency has the effect of increasing the local detuning $\delta_a(z_t)$ to preserve the same population. This manifests itself as a ``self-selection'' of the trapping position and an increasing distance from the barrier,
\be \frac{z_t-z_b}{z_b}\propto\frac{1}{\Delta_p}\left(\tilde{\Omega}^2-1\right)^{1/2}, \ee
where we have defined the dimensionless Rabi frequency $\tilde{\Omega}=\Omega/\Omegamin$.

The photon scattering rate at the trap position, $R_{\footnotesize\textrm{sc}}=\Gamma(z_t)\sigma_{ee}(z_t)$, leads to lifetime limitations of our trapping scheme. Fixing other parameters, the detuning $\Delta_p$ from the plasmon resonance can be optimized to yield the minimum scattering rate, $R_{\footnotesize\textrm{sc,min}}\sim\frac{\Gamma_0}{\sqrt{Q(k_0 z_t)^{3/2}}}$. Thus, absent fundamental limitations to the magnitude of $Q$, the photon scattering rate can in principle be suppressed to an arbitrary degree. This implies that an ``infinitesimal'' violation of the assumptions underlying the no-go theorem can in fact allow for trapping.

In a conventional optical dipole trap, the potential $U=-\alpha(\omega_L)|E(\bfr,\omega_L)|^2/2$ seen by an atom depends on the local field intensity and is proportional to a spatially-constant atomic polarizability evaluated at the laser frequency. In contrast, here we exploit a novel mechanism arising from strong spatial variations in the polarizability itself, induced through the vacuum level shifts.  This creates a back-action effect analogous to that producing a dynamical ``optical spring'' in opto-mechanical systems~\cite{corbitt_all-optical_2007}, and enables extremely steep trap barriers around $z_b$~(as in Fig.~\ref{fig:schematic}c). In particular, the repulsive force is given by $F\sim-\frac{d\omega_e}{dz}\sigma_{ee}(z)$, where $\frac{d\omega_e}{dz}$ is engineered to be large. However, the excited state population $\sigma_{ee}(z)\sim\frac{\Omega^2}{4\delta_a(z)^2}$ itself varies strongly with position through the spatially-dependent detuning~(for simplicity, here we omit the varying linewidth $\Gamma(z)$). This provides an overall repulsive force that scales like $(\frac{d\omega_e}{dz})^2$,
\be F\sim-\left(\frac{d\omega_e(z_t)}{dz}\right)^2\frac{2(z-z_t)\sigma_{ee}(z_t)}{\delta_a(z_t)}. \ee
%

\begin{figure}[t]
\begin{center}
\includegraphics[width=15cm]{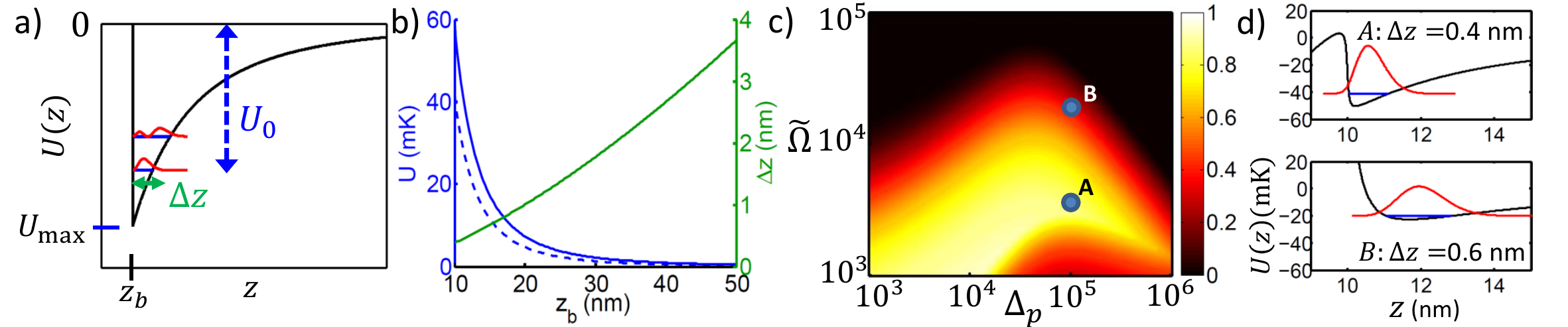}
\end{center}
\caption{\textbf{Trap properties.} a) Ideal limit of trapping potential, corresponding to the pure ground-state potential $\hbar\delta\omega_g(z)$ and an infinite barrier at $z=z_b$. The maximum depth of the potential is denoted by $U_{\footnotesize\textrm{max}}\equiv\hbar|\delta\omega_g(z_b)|$. The quantum wave-function amplitudes $|\psi(z)|^2$~(in arbitrary units) for the two lowest eigenstates are illustrated in red, while the binding energy $U_0$ and position uncertainty $\Delta z$ of the ground state are depicted as well. b) Properties of the ideal trapping potential versus barrier position $z_b$, for the case of a Cesium atom. The blue solid and dashed curves give the potential depth at the barrier position, $U_{\footnotesize\textrm{max}}$, and the ground-state binding energy $U_0$, respectively. The green curve depicts the ground-state uncertainty $\Delta z$. c) For the non-ideal trapping potential, the depth of the classical potential $U_{\footnotesize\textrm{depth}}$ is numerically calculated. Here we plot it normalized by the maximum possible value $U_{\footnotesize\textrm{max}}\equiv\hbar|\delta\omega_g(z_b)|$, versus detuning parameter $\Delta_p$ and normalized Rabi freuqency $\tilde{\Omega}\equiv\Omega/\Omegamin$. d) The trapping potential, ground-state binding energy $U_0$, and wave-function amplitude $|\psi(z)|^2$ for the parameter sets A,B, as illustrated in c). The color coding is the same as in a).\label{fig:properties}}
\end{figure}

Given the possible steepness of this barrier, an ``ideal'' limit for the overall potential is shown in Fig.~\ref{fig:properties}a. The attractive part consists of the pure ground-state potential and arises from quantum fluctuations, while an infinite ``hard wall'' is created at $z=z_b$ due to the combination of strong resonant excited-state shifts~(originating from the interaction of the atom with a ``classical'' image dipole) and opto-mechanical back-action. This idealized model enables one to understand the best possible scaling of the trap properties. In particular, the trap depth $U_{\footnotesize\textrm{depth}}$~(the energy required for a classical particle to become unbound) cannot exceed the value of the ground-state potential at $z_b$, $U_{\footnotesize\textrm{max}}\equiv\hbar |\delta\omega_g(z_b)|$~(\textit{e.g.}, $U_{\footnotesize\textrm{max}}\sim 60$~mK when $z_b=10$~nm for a Cesium atom). This trapping potential can be easily quantized numerically. In Fig.~\ref{fig:properties}b, we plot the quantum binding energy $U_0$~(where $0<U_0<U_{\footnotesize\textrm{depth}}\leq U_{\footnotesize\textrm{max}}$) and position uncertainty $\Delta z$ for the motional ground-state wave function. For the numerical results in Figs.~\ref{fig:properties} and~\ref{fig:lifetime}, we use atomic properties corresponding to Cesium~($\lambda_0=852$~nm, $\Gamma_0/2\pi=5.2$~MHz, recoil frequency $\omega_r/\Gamma_0=4\times 10^{-4}$). The strength of the vacuum potentials is reflected in the strong confinement of the wave-functions~(\textit{e.g.}, $\Delta z<4$~nm for distances $z_b<50$~nm), which are an order of magnitude smaller than in conventional optical traps. To contrast our results for Cesium with those of a less massive atom, we have considered Lithium as well. Results for Li similar to those for Cs in Fig.~\ref{fig:properties}a,b show that at $z_b=10$~nm, the potential depth is smaller by roughly a factor of two, while the confinement $\Delta z$ sees roughly a five-fold increase. Generally, within the WKB approximation and to lowest order in $\Delta z$, the ground-state localization scales like $k_0 \Delta z\sim \left(\omega_r(k_0 z_b)^4/\Gamma_0\right)^{1/3}$, while the binding energy $U_0 \sim U_{\footnotesize\textrm{max}}(1-3\Delta z/z_b)$ can be nearly the entire classical potential depth.

This ideal model describes qualitatively well the actual potentials over a large parameter regime. In Fig.~\ref{fig:properties}c, we plot the ratio of the numerically obtained trap depth to the theoretical maximum, $U_{\footnotesize\textrm{depth}}/U_{\footnotesize\textrm{max}}$, as functions of $\Delta_p$ and dimensionless driving amplitude $\tilde{\Omega}$. Here, we have chosen a quality factor of $Q=10^7$ and a barrier position of $z_b=10$~nm. Although we have provided near-field approximations for the energy shifts above, the plots in Figs.~\ref{fig:properties}c,d and Fig.~\ref{fig:lifetime} are calculated using the full Green's functions. A trap depth comparable to $U_{\footnotesize\textrm{max}}$ is achievable over a large range of values. For low values of $\tilde{\Omega}$, the trap depth is weakened due to escape over the potential barrier, while for large values, the trap position becomes shallow as $z_t$ is pulled away from $z_b$~(as discussed previously). Two representative trap potentials, along with the ground-state wave functions and confinement $\Delta z$ are illustrated in Fig.~\ref{fig:properties}d.

\section{Trap lifetime}

We have identified four sources of motional heating -- recoil heating, non-adiabatic motion of the atom within the trap, fluctuations of the atomic coherence, and tunneling over the finite-height potential barrier -- that limit the trap lifetime in absence of external cooling mechanisms. Tunneling is exponentially suppressed with barrier height and thus is easily suppressed with increasing Rabi frequency. This calculation can be found in the Appendix, while we discuss the more important mechanisms here.

As in conventional optical trapping, the random momentum recoil imparted on the atom by scattered photons yields an increase in motional energy at the rate $dE/dt=\frac{(\hbar k_{\footnotesize\textrm{eff}})^2}{2m}R_{\footnotesize\textrm{sc}}$. Two qualitative differences emerge relative to free space, however. First, the photon scattering rate $R_{\footnotesize\textrm{sc}}=\Gamma(\bfr)\sigma_{ee}\sim \frac{\Gamma_0\sigma_{ee}}{16(k_0 z)^3}\frac{Q}{\Delta_p^2}$ is modified in the presence of dielectric surfaces. Second, the effective momentum $\hbar k_{\footnotesize\textrm{eff}}$ imparted by a photon is enhanced compared to the free-space momentum. This momentum scales at close distances like $k_{\footnotesize\textrm{eff}}\sim\sqrt{3}/z$~(see Appendix), due to emission into high-wavevector photons in the dielectric. Starting from the motional ground state, this heating causes the atom to become unbound over a time scale $\tau_r\sim(k_0 z_t)^2\Delta_p/(3\omega_r)$. It should be noted that the heating rate $dE/dt$ does not depend on the atom confinement~(\textit{i.e.}, whether we operate in the Lamb-Dicke limit), as the suppression of motional jumps in tight traps is compensated by the increase in energy gained per jump~\cite{cirac_laser_1992}.

Our derivation of the atomic potential assumes that the atomic coherence adjusts rapidly to the local trap properties, $d\sigma_{ge}/dt\approx 0$, which results in a purely conservative potential. As we now show, non-adiabatic corrections result in a momentum-dependent force, $dp/dt\propto p$~(anti-damping). We can solve Eq.~(\ref{eq:Bloch}) for the atomic motion perturbatively by writing $\sigma_{ge}(t)=\sigma_{ge}^{(0)}(z(t))+\sigma_{ge}^{(1)}(t)$, where $\sigma_{ge}^{(0)}$ is the adiabatic solution. It depends only on the instantaneous position and is given by $\sigma_{ge}^{(0)}\equiv\frac{\Omega}{2\delta_c(z)}$ in the weak saturation limit. Here, $\delta_c(z)=\delta_a(z)+i\Gamma(z)/2$ is the complex detuning. Substituting this solution into Eq.~(\ref{eq:Bloch}) yields a velocity-dependent correction to the coherence, $\sigma_{ge}^{(1)}=(i\delta_c)^{-1}\frac{d\sigma_{ge}^{(0)}}{dt}=-\frac{\Omega}{2i\delta_c^3}\frac{d\delta_c}{dz}\frac{dz}{dt}$. This in turn yields a new contribution to the force, $F^{(1)}=\beta p$, where the anti-damping rate is given by $\beta\approx-\omega_r \frac{d\omega_e}{dz}\frac{\sigma_{ee}}{2k_0^2|\delta_c|^4}\left[(\Gamma^2-4\delta_a^2)\frac{d\Gamma}{dz}+8\delta_a\Gamma\frac{d\delta_a}{dz}\right]$ evaluated at the trap position $z_t$. Intuitively, the atom is anti-damped~($\beta>0$) because the laser frequency is blue-detuned relative to the atomic frequency at $z_t$, giving rise to preferential Stokes scattering over anti-Stokes. Aside from the term proportional to $d\Gamma/dz$, the heating rate is exactly analogous to that occurring in an opto-mechanical system in the blue-detuned regime~\cite{corbitt_all-optical_2007}. We define a characteristic time $\tau_{\footnotesize\textrm{ad}}$ needed for the energy increase produced by anti-damping to equal the ground-state binding energy $U_0$.

Thus far, we have focused on the motion of the atomic dressed state, in which a trap emerges due to weak mixing with the excited state potential. Following a spontaneous emission event, however, the atom returns to the ground state and sees the pure ground-state potential for a transient time $t_{\footnotesize\textrm{trans}}\sim|\delta_c(z_t)|^{-1}$ until the atomic coherence is restored, during which the atomic wave-packet accelerates toward the surface and gains energy. While a simulation of the full dynamics of the atom~(including the spatial and internal degrees of freedom) is quite challenging, here we estimate the heating rate from these transient processes. In particular, we first find the spatial propagator under the pure ground-state potential, $U(z_f,z_i,t_f,t_i)$~(see Appendix). Then, assuming that a spontaneous emission event occurs at $t_i=0$ with the atom initially in the motional ground-state $\psi(z_i)$, the new wave-function that emerges once the coherence equilibrates is given by $\psi(z_f)\sim\int\,dz_i U(z_f,z_i,t_{\footnotesize\textrm{trans}},0)\psi(z_i)$. The probability of atom loss $P_{\footnotesize\textrm{trans}}$ per spontaneous emission event is estimated from the overlap between the initial and final states, $P_{\footnotesize\textrm{trans}}=1-\left|\langle z_f | z_i \rangle \right|^2$, with a corresponding trap lifetime of $\tau_{\footnotesize\textrm{trans}}^{-1}=P_{\footnotesize\textrm{trans}}R_{\footnotesize\textrm{sc}}$.

\begin{figure}[t]
\begin{center}
\includegraphics[width=8cm]{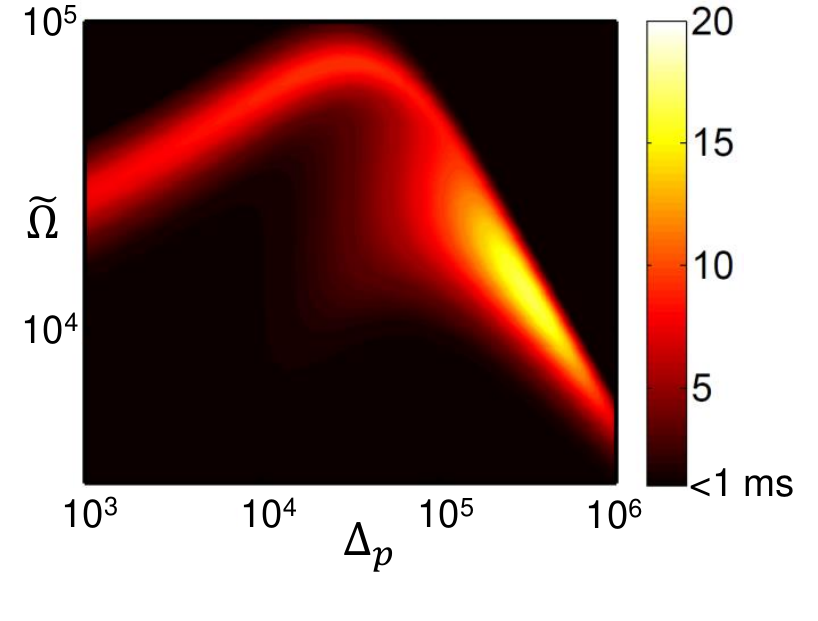}
\end{center}
\caption{\textbf{Trap lifetime.} Trap lifetime $\tau_{\footnotesize\textrm{total}}$ versus dimensionless Rabi frequency $\tilde{\Omega}=\Omega/\Omegamin$ and detuning $\Delta_p$. Here we have assumed a plamson resonance quality factor of $Q=10^7$ and atomic properties corresponding to Cesium. \label{fig:lifetime}}
\end{figure}

In Fig.~\ref{fig:lifetime}, we plot the overall trap lifetime, which is obtained by adding the individual lifetimes in parallel, $\tau_{\footnotesize\textrm{total}}^{-1}=(\tau_{\footnotesize\textrm{r}}^{-1}+\tau_{\footnotesize\textrm{ad}}^{-1}+\tau_{\footnotesize\textrm{trans}}^{-1}+\tau_{\footnotesize\textrm{tunnel}}^{-1})$. The overall trap lifetime exhibits a complicated dependence on Rabi frequency $\Omega$ and detuning $\Delta_p$ due to the different scalings of the constituent heating mechanisms. As an example, however, for external parameters $\Omega/\Omegamin=1.5\times 10^4$ and $\Delta_p=2.5\times 10^5$ for atomic Cs, one achieves a lifetime of $\tau_{\footnotesize\textrm{total}}\approx 15$~ms, ground-state confinement $\Delta z\approx 1$~nm, trap depth $U_{\footnotesize\textrm{depth}}\approx 10$~mK, and trap distance $z_t\approx 15$~nm.

\section{Outlook}

We have described a protocol for an atomic trap based upon engineered vacuum forces, and have analyzed in detail a model case of one-dimensional trapping near a Drude material, where analytical results are possible to obtain. The Drude response approximates well a number of metals, such as silver and gold, in the optical domain~\cite{johnson_optical_1972}. In practice, however, these metals are limited by their low quality factors~($Q<10^2$), and the thermal fluctuations of such conducting materials can give rise to strong noise-induced trap heating and decoherence~\cite{henkel_loss_1999}. On the other hand, photonic crystal structures enable engineering of resonances through geometry, and quality factors approaching $Q\sim 10^7$ have been observed~\cite{taguchi_statistical_2011}. While the field profiles of such structures are more complex, we anticipate that our trapping protocol could be quite generally applied, and the resulting trap behavior would qualitatively remain the same. In particular, the scaling of the engineered excited-state shift~($\delta\omega_e\propto Q/\Delta_p$) and dissipation~($\Gamma\propto Q/\Delta_p^2$) are generic to coupling between an atom and resonator, and only the spatial dependence is determined by the field profiles. Photonic crystals should thus offer great flexibility in tailoring the properties of vacuum traps~(such as dimensionality), which we plan to investigate in detail in future work. We note that related efforts have already been made to design realistic hybrid atom traps in photonic crystal structures, where a combination of vacuum forces and optical forces are required to stabilize the trap in all dimensions~\cite{hung_trapped_2013}. In addition, while we have focused on trapping of atoms weakly dressed by an external laser, it should also be possible to achieve laser-free traps by using atoms pumped into metastable electronic levels and engineering resonant shifts.

The ability to create atomic traps with parameter sets~(such as depth, confinement, and proximity to surfaces) that are not possible in conventional traps~(whether optical, electrical, or magnetic) should lead to a number of intriguing applications. For example, it has been proposed that the trapping of atoms near dielectric surfaces with nano-scale features~(such as a sub-wavelength lattice) would facilitate large interaction strengths and long-range interactions for ultracold atoms~\cite{gullans_nanoplasmonic_2012}. However, the major limitation of optical trapping is the divergent intensity needed to overcome the ground-state vacuum forces of these structures. The use of resonantly-enhanced excited state repulsion to create a trap could significantly reduce the power requirements and enable smaller lattice constants. Furthermore, traps with localization on the nanometer scale can induce inter-atomic forces that are of comparable strength to molecular van der Waals forces, which enables novel opportunities to control ultracold atomic collisions~\cite{bolda_effective-scattering-length_2002} and realize exotic interactions such as $p$-wave scattering of fermions~\cite{julia-diaz_engineering_2013}. Atoms trapped near surfaces could also act as exquisite nanoscale probes of surface physics, including the precision measurement of vacuum forces themselves.

The authors thank N. Stern and O. Painter for helpful discussions. DEC acknowledges support from Fundaci\'{o} Privada Cellex Barcelona. KS was funded by the NSF Physics Frontier Center at the JQI. JMT acknowledges funding from the NSF Physics Frontier Center at the JQI and the US Army Research Office MURI award W911NF0910406. HJK acknowledges funding from the IQIM, an NSF Physics Frontier Center with support of the Moore Foundation, by the AFOSR QuMPASS MURI, by the DoD NSSEFF program, and by NSF PHY-1205729.

\appendix

\section{Electromagnetic field quantization}

In this section we briefly describe the quantization scheme of the electromagnetic field in the presence of arbitrary linear dielectric media, which follows that of Ref.~\cite{buhmann_dispersion_2007}. The free field is characterized by a set of bosonic modes with frequencies $\omega$ and annihilation operators $\hat{a}_j(\bfr,\omega)$~($j=x,y,z$), with a corresponding Hamiltonian
\be H_f = \sum_j\int d\bfr \int\mathrm{d}\omega \, \omega  \hat{a}_j^\dagger(\bfr,\omega)\hat{a}_j(\bfr,\omega). \ee
The bosonic operators satisfy canonical commutation relations $[\hat{a}_j(\bfr,\omega),\hat{a}_k^\dagger(\bfr',\omega')]=\delta_{jk}\delta(\bfr-\bfr')\delta(\omega-\omega')$ and are physically associated with noise polarization sources in the material. The electric field operator at frequency $\omega$ is driven by the sources at the same frequency and is given in the Coulomb gauge by~\cite{buhmann_dispersion_2007}
\be E_i(\bfr,\omega)= i\mu_0 \omega^2 \int d\bfr' G_{ij}(\bfr,\bfr',\omega)\sqrt{\frac{\hbar\epsilon_0}{\pi}\textrm{Im}\,\epsilon(\bfr',\omega)}\hat{a}_j(\bfr',\omega).\label{eq:E} \ee
Here and in the following, it is assumed that all repeated vector or tensor indices are summed over, \textit{e.g.}, $G_{ij}\hat{a}_j=\sum_{j}G_{ij}\hat{a}_j$. $\epsilon(\bfr,\omega)$ is the frequency-dependent permittivity at position $\bfr$, and $G_{ij}$ is the dyadic Green's function which is the (gauge-invariant) solution to
\be \left[\left(\nabla\times\nabla\times\right)-\frac{\omega^2}{c^2}\epsilon(\bfr,\omega)\right]G(\bfr,\bfr',\omega)=\delta(\bfr-\bfr')\otimes I.\label{eq:Greeneq} \ee
The total electric field is ${\bf E}(\bfr)=\int d\omega\,{\bf E}(\bfr,\omega)+h.c.$ Similarly, the magnetic field is ${\bf B}(\bfr)=\int d\omega\,\frac{1}{i\omega}\nabla\times{\bf E}(\bfr,\omega)+h.c$. It can be verified that these expressions preserve the same field commutation relations as in free-space field quantization.

In the following, we consider a two-level atom at position $\bfr_a$ and with ground and excited states $\ket{g}$,$\ket{e}$ interacting with the electromagnetic field within the electric dipole approximation. The interaction Hamiltonian is given by
\bea H_{\footnotesize\textrm{af}} & = & -{\bf d}\cdot{\bf E}(\bfr_a) \\
& = & -\wp_i(\sigma_{eg}E_i(\bfr_a)+h.c.),\label{eq:Haf} \eea
where $\wp_i$ is the atomic dipole matrix element.

\section{Green's function for planar interface}\label{sec:recoil}

Our system of interest consists of an atom in a region of vacuum with nearby dielectric objects. In general, for $\bfr,\bfr'$ in the vacuum region, one can express the Green's function as a sum $G=G_{\footnotesize\textrm{free}}+G_{\footnotesize\textrm{sc}}$, where the free component is the solution in uniform vacuum~($\epsilon=1$) and the scattered component physically describes waves scattered from the dielectric objects.

In a planar geometry such as that considered in the main text, the Green's function can be evaluated exactly using a plane wave expansion. Specifically, suppose that we have an interface with vacuum in the region $z>0$ and a dielectric with permittivity $\epsilon(\omega)$ in the region $z<0$. The free and scattered components of the dyadic Green's function in the region $0\leq z \leq z'$ are then given by
\bea G_{\footnotesize\textrm{free}} & = & -\frac{c^2}{4\pi^2\omega^2}\int d\bfk_{\parallel} e^{i\bfk_{\parallel}\cdot(\boldrho-\boldrho')}\left[\delta(z-z')\hat{z}\hat{z}+\frac{i}{2\kperp}e^{i\kperp(z'-z)}\bfk_{0}\times\bfk_{0}\times\right],\label{eq:Gfree} \\ G_{\footnotesize\textrm{sc}} & = & \frac{ic^2}{8\pi^2\omega^2}\int d\bfk_{\parallel}e^{i\bfk_{\parallel}\cdot(\boldrho-\boldrho')} e^{i\kperp(z+z')} \frac{1}{\kperp} \left[r_p(\kp\hat{z}-\kperp\hat{k}_{\parallel})(\kp\hat{z}+\kperp\hat{k}_{\parallel})+r_s \frac{\omega^2}{c^2}(\hat{z}\times\hat{k}_{\parallel})(\hat{z}\times\hat{k}_{\parallel})\right].\label{eq:Gsc} \eea
Here $\bfk_{\parallel}=(k_x,k_y)$ and $\boldrho=(x,y)$ are the wavevector and position along the direction parallel to the interface, $\bfk_0=\bfk_{\parallel}-\kperp\hat{z}$, $\kp^2+\kperp^2=(\omega/c)^2$, and $r_s$ and $r_p$ are the Fresnel reflection coefficients for $s$ and $p$ polarized waves, respectively.

\section{Ground-state interactions}\label{sec:ground}

The interaction Hamiltonian $H_{\footnotesize\textrm{af}}$ couples the ground state of the bare system~$\ket{g,0}$~(consisting of the atomic ground state and vacuum) to states $\ket{e,1_{\bfr,\omega,j}}\equiv\hat{a}^{\dagger}_j(\bfr,\omega)\ket{e,0}$. Within second-order perturbation theory, this interaction induces a shift of the bare ground state energy by an amount
\be \delta\omega_g(\bfr_a)=-\frac{1}{\hbar^2}\sum_{j}\int d\bfr \int d\omega \frac{\bra{e,1_{\bfr,\omega,j}} H_{\footnotesize\textrm{af}}\ket{g,0}|^2}{\omega+\omega_0}, \ee
where $\omega_0$ is the bare resonance frequency of the atom. Substituting Eqs.~(\ref{eq:E}) and~(\ref{eq:Haf}) into the expression above, and using analyticity of the integrand to rotate the frequency integral onto the positive imaginary axis $\omega=iu$, one finds~\cite{buhmann_dispersion_2007}
\be \delta\omega_g(\bfr_a) = \frac{\mu_0}{\hbar\pi}\int_0^{\infty} du \frac{u^2\omega_0}{\omega_0^2+u^2}\wp_i G_{\footnotesize\textrm{sc},ij}(\bfr_a,\bfr_a,iu) \wp_j.\label{eq:omegag} \ee
Here we have only included the scattered component of the Green's function, as the free component gives a contribution that is independent of position. While we have considered a two-level atom for simplicity, a more realistic model of an isotropic atom consists of multiple excited states and equal polarizabilities in each direction~(\textit{i.e.}, the ground state can emit virtual photons of any polarization with equal strength). Since the contribution to the ground-state shift from each excited state transition is additive, Eq.~(\ref{eq:omegag}) can be modified to account for this by setting $\wp_i=\wp_0$ for all $i$ and associating $\omega_0^3 \wp_0^2/(3\pi\epsilon_0\hbar c^3)=\Gamma_0$ with the free-space spontaneous emission rate of any excited state. This yields Eq.~(1) in the main text.

\section{Excited-state interactions}

In this section, because we want to focus exclusively on the effect of vacuum interactions with the excited state, it is sufficient to include only energy-conserving terms in the atom-field interaction Hamiltonian $H_{\footnotesize\textrm{af}}$~(\textit{i.e.}, allowing a transition from the excited to ground state accompanied by creation of a photon from vacuum). We also want to consider the specific case where the atom is weakly driven by a classical field of frequency $\omega_L$, which motivates working in a rotating frame where the free atomic evolution is given by $H_a=-\delta\sigma_{ee}$ and $\delta=\omega_L-\omega_0$ is the detuning between the laser frequency and atomic resonance frequency $\omega_0$. As will be seen, the inclusion of the laser frequency in the free Hamiltonian causes the effective excited state shift and decay rates to depend on the Green's function evaluated at $\omega_L$ once the vacuum modes are eliminated. Physically, this describes the effect of the weakly driven atom~(which primarily acts as a Rayleigh scatterer) interacting with its own Rayleigh-scattered field, which can be altered due to the presence of nearby dielectric surfaces.

The standard technique to derive the excited-state shift and decay rate follows that of Appendix~\ref{sec:ground} for the ground state and uses time-independent second-order perturbation theory~\cite{buhmann_dispersion_2007}. In this approach, however, one cannot find the jump operator associated with photon emission, which is needed in order to calculate the effect of emission on atomic motion~(\textit{e.g.}, recoil heating). To rectify this, we employ an alternative approach based on deriving an atomic master equation within the Born-Markov approximation~\cite{meystre_elements_2007}. In particular, the equation of motion for the reduced atomic density matrix $\rho_a$ is given by
\be
\dot{\rho}_a =\frac{-1}{\hbar^2}\mathrm{Tr}_f\int_0^\infty\mathrm{d}\tau \left[\tilde{H}_{\footnotesize\textrm{af}}(t),\left[\tilde{H}_{\footnotesize\textrm{af}}(t-\tau),\rho_a\otimes\ket{0}\bra{0}\right]\right],\label{eq:rhodot} \ee
where $\tilde{H}_{\footnotesize\textrm{af}}$ is the interaction Hamiltonian in the interaction picture~(with respect to the free Hamiltonian $H_0=H_a+H_f$) and $\ket{0}$ is the electromagnetic vacuum state.  Any exponential of time appearing in the integral can be evaluated using the relation $\int_{0}^{\infty}d\tau\,e^{i\omega\tau}=\pi\delta(\omega)+i\mathcal{P}\frac{1}{\omega}$. Using the field expansion of Eq.~(\ref{eq:E}) in $\tilde{H}_{\footnotesize\textrm{af}}$ and following some manipulation, Eq.~(\ref{eq:rhodot}) can be written in the form
\be \dot{\rho}_a=-(i/\hbar)[H_{\footnotesize\textrm{eff}},\rho_a]+\mathcal{L}[\rho_a]. \ee

The effective Hamiltonian describes coherent interactions between the atom and vacuum field and takes the form
\be H_{\footnotesize\textrm{eff}}=\hbar\delta\omega_e(\bfr_a)\sigma_{ee}, \ee
which can be interpreted as a position-dependent energy shift of the excited state. A realistic model of an atom allows for the possibility for an excited state to emit photons of different polarizations and transition into different states in a ground-state manifold. The different Clebsch-Gordan coefficients of these transitions, along with the symmetry breaking introduced by the nearby dielectric structure, imply that different excited states can experience different shifts absent some special engineering. To arrive at a simpler two-level atomic model that captures the essential physics of our trapping mechanism, we assume that the excited state is equally coupled to all polarizations~(but still with total emission rate $\Gamma_0$), such that one should average over all dipole orientations. In this case, the excited-state shift is given by
\be \delta\omega_e(\bfr_a)=-\frac{c\Gamma_0}{\omega_0^2}\int_{0}^{\infty}du\frac{u^2}{\omega_0^2+u^2}\textrm{Tr}\,G_{\footnotesize\textrm{sc}}(\bfr_a,\bfr_a,iu)-\frac{\Gamma_0 \pi c}{\omega_L}\textrm{Tr}\,\textrm{Re}\,G_{\footnotesize\textrm{sc}}(\bfr_a,\bfr_a,\omega_L),\label{eq:omegaeappendix} \ee
as stated in the main text.

The Liouvillian term $\mathcal{L}[\rho_a]$ can be written in a Lindblad form
\be \mathcal{L}[\rho_a]=\sum_{j}\int d\bfk \, \mathcal{O}_{\bfk j}\rho_a \mathcal{O}_{\bfk j}^{\dagger}-\frac{1}{2}(\mathcal{O}_{\bfk j}^{\dagger}\mathcal{O}_{\bfk j}\rho_a+\rho_{a}\mathcal{O}_{\bfk j}^{\dagger}\mathcal{O}_{\bfk j}). \ee
The jump operators take the form
\be
\mathcal{O}_{\bfk j}(\bfr_a)= \sqrt{\frac{2\epsilon_0\mu_0^2\omega_L^4}{\hbar}}\int d\bfr\frac{e^{i\bfk \cdot \bfr}}{(2\pi)^{3/2}}\sqrt{\mathrm{Im}\epsilon(\bfr,\omega_L)} \wp_i G^\ast_{ij}(\bfr_a,\bfr,\omega_L)\sigma_{ge}.\label{eq:Ojump} \ee
The excited-state spontaneous emission rate is in turn given through the relation $\sum_{j} \int d\bfk\,\mathcal{O}_{\bfk j}^{\dagger}\mathcal{O}_{\bfk j}=\Gamma(\bfr_a)\sigma_{ee}$, or
\be \Gamma(\bfr_a)
=\Gamma_0+\frac{2\Gamma_0 \pi c}{\omega_L}\mathrm{Tr}\,\mathrm{Im}\,G_{\footnotesize\textrm{sc}}(\bfr_a,\bfr_a,\omega_L).\label{eq:Gammaappendix} \ee
Eq.~(\ref{eq:Gammaappendix}) again averages over dipole orientations and reproduces the expression for the emission rate provided in the main text. To derive this, we have used the identity
\be \frac{\omega^2}{c^2}\int d\bfr''\,\left(\mathrm{Im}\,\epsilon(\bfr'',\omega)\right) G_{ij}(\bfr,\bfr'',\omega)G_{jk}^{\ast}(\bfr'',\bfr',\omega)=\mathrm{Im}\,G_{ik}(\bfr,\bfr',\omega). \ee

The expressions for the excited-state emission rate $\Gamma(\bfr_a)$ and level shift $\delta\omega_e(\bfr_a)$ agree with previous derivations~\cite{buhmann_dispersion_2007}, if the laser frequency $\omega_L$ is replaced by the atomic resonance frequency $\omega_0$. The appearance of the laser frequency here is associated with the restriction to the weak-driving limit, which was not considered in earlier derivations. The main new result obtained here, however, is an explicit expression for the jump operator associated with photon emission, which we will use to calculate photon recoil heating in Appendix~\ref{sec:recoil}.

\section{Motional heating}

Here we derive in more detail the equations relating to heating of the atomic motion around the minimum of the vacuum force trap.

\subsection{Recoil heating}\label{sec:recoil}

Recoil heating can be calculated by treating the atomic position $\bfr_a$ in Eq.~(\ref{eq:Ojump}) as an operator. For notational simplicity, we shall consider recoil heating only along one direction, say $z$, although our results can be easily generalized. In the main text, $z$ corresponds to the direction in which the atom is trapped~(normal to the dielectric surface). The increase in momentum uncertainty due to photon scattering is given by
\be \frac{d}{dt}\avg{p_z^2} = \mathrm{Tr}(p_z^2 \dot{\rho}_a). \ee
Here we will focus solely on the Liouvillian term in the density matrix evolution~(\textit{i.e.}, momentum incurred from spontaneous emission). It can be shown that the Liouvillian term produces no net force, \textit{i.e.}, $\frac{d}{dt}\avg{p_z}=0$. Using the fact that $[p_z,f(z_a)]=-i\hbar\frac{\partial f}{\partial z_a}$ for any function $f$, one can derive the following general expression for the recoil heating rate in terms of Green's functions,
\be \frac{d}{dt}\avg{p_z^2}=\hbar\mu_0\omega_L^2 \wp_i \wp_j \left(\partial_1^2+2\partial_1\partial_2-\partial_2^2\right)\avg{\mathrm{Im}\,G_{ij}(z_a,z_a,\omega_L)\sigma_{ee}}.\label{eq:recoil} \ee
Here, the derivatives $\partial_1$ and $\partial_2$ act on the first and second spatial arguments of the Green's functions, respectively, and we have suppressed the atomic spatial variables in the directions that are not of interest. We further assume that the internal and spatial degrees of freedom can be de-correlated, $\avg{G_{ij}\sigma_{ee}}\approx \avg{G_{ij}}\avg{\sigma_{ee}}$. Formally, we can write the above equation in the convenient form
\be \frac{d}{dt}\avg{p_z^2}=(\hbar k_{\footnotesize\textrm{eff}})^2 \Gamma(\bfr_a) P_e, \ee
where $P_e=\avg{\sigma_{ee}}$ is the excited-state population. This form is intuitive as it describes the momentum increase arising from a random walk process, where the atom experiences random momentum kicks of size $\pm\hbar k_{\footnotesize\textrm{eff}}$ at a rate $\Gamma(\bfr_a) P_e$~(\textit{i.e.}, the photon scattering rate). $k_{\footnotesize\textrm{eff}}$ is a derived quantity that thus characterizes the effective momentum associated with scattered photons in the vicinity of a dielectric surface.

As a simple example, we can consider an atom in free space polarized parallel to the surface~(say along $x$). Evaluating Eq.~(\ref{eq:recoil}) with the free-space Green's function $G_{\footnotesize\textrm{free},xx}$ from Eq.~(\ref{eq:Gfree}) yields $k_{\footnotesize\textrm{eff}}=\sqrt{2/5}(\omega_L/c)$~(and $\Gamma(\bfr_a)=\Gamma_0$), recovering the known result~\cite{cirac_laser_1992}. This recoil momentum is smaller than the full momentum $\omega_L/c$ of the scattered photon because the atom emits in a dipole pattern, but only the projection of the photon momentum along $z$ contributes to motional heating in that direction.

As a more complicated example, we can consider the recoil heating near a surface in the trapping scheme considered in the main text. For the Drude model and at small distances between the atom and surface, we find that the scattered component of the Green's function dominates the recoil heating, leading to the asymptotic expansion of $k_{\footnotesize\textrm{eff}}\approx\sqrt{3}/z_a$ given in the main text. In all of our numerical calculations, the full Green's function is used rather than asymptotic results.

\subsection{Transient heating}

As described in the main text, following a spontaneous emission event, the spatial wave-function of the atom is temporarily untrapped and evolves under the pure ground-state potential for a characteristic time $t_{\footnotesize\textrm{trans}}$ before the internal dynamics equilibrates. Here we discuss how to calculate an approximate propagator $U(z_f,z_i,t_f,t_i)$ for the wave-function evolving under the ground-state potential.

The problem significantly simplifies if the ground-state potential $U_g(z)=\hbar\omega_g(z)$ is linearized around the atom trapping position $z=z_t$,
\be H\approx \frac{p_z^2}{2m}+U_g(z_t)+mg(z-z_t), \ee
where $mg=U_{g}'(z_t)$. This linearization is justified in our regime of interest where the atom is tightly trapped and the wave-function undergoes little evolution in the time $t_{\footnotesize\textrm{trans}}$. For simplicity, here we will ignore the constant energy and position offsets and consider the simplified Hamiltonian $H=\frac{p_z^2}{2m}+mgz$, which corresponds to a mass accelerating under a constant gravitational force. This observation enables one to solve for the propagator by utilizing the equivalence principle, that the system of interest is equivalent to that of a free particle viewed in an accelerating frame. One finds that the propagator is given by
\be U(z_f,z_i,t,0)=U_{\footnotesize\textrm{free}}(z_f,z_i,t)\exp\left(-\frac{imgt(z_f+z_i)}{2\hbar}-\frac{img^2 t^3}{24\hbar}\right),\label{eq:prop} \ee
where $U_{\footnotesize\textrm{free}}(z_f,z_i,t)=\sqrt{\frac{m}{2\pi i \hbar t}}\exp\left(\frac{im(z_f-z_i)^2}{2\hbar t}\right)$ is the well-known propagator for a free particle in an inertial frame~\cite{sakurai_modern_1994}.

In our simulations, the ground state wave-function for an arbitrary trapping potential is obtained numerically on a lattice, and so the propagator must be spatially discretized as well. In particular, while the exponential term in Eq.~(\ref{eq:prop}) is well-behaved on a lattice, the free propagation $U_{\footnotesize\textrm{free}}$ is not at short times. We instead replace $U_{\footnotesize\textrm{free}}$ with the propagator $U_{mn}(t)$ for the free discrete Schrodinger equation,
\be \frac{\partial\psi_n}{\partial t}=i\beta(\psi_{n+1}(t)-2\psi_n(t)+\psi_{n-1}(t)), \ee
where $\beta=\hbar/2ma^2$, $n$ is the lattice site index, and $a$ is the lattice constant. The propagator for this equation (defined as the matrix satisfying $\psi_m(t)=D_{mn}\psi_n(0)$) is given by
\be D_{mn}(t)=i^{m-n}J_{m-n}(2\beta t)e^{-2i\beta t}, \ee
where $J_n$ is the $n$-th order Bessel function.

\subsection{Tunneling}

\begin{figure}[t]
\begin{center}
\includegraphics[width=7cm]{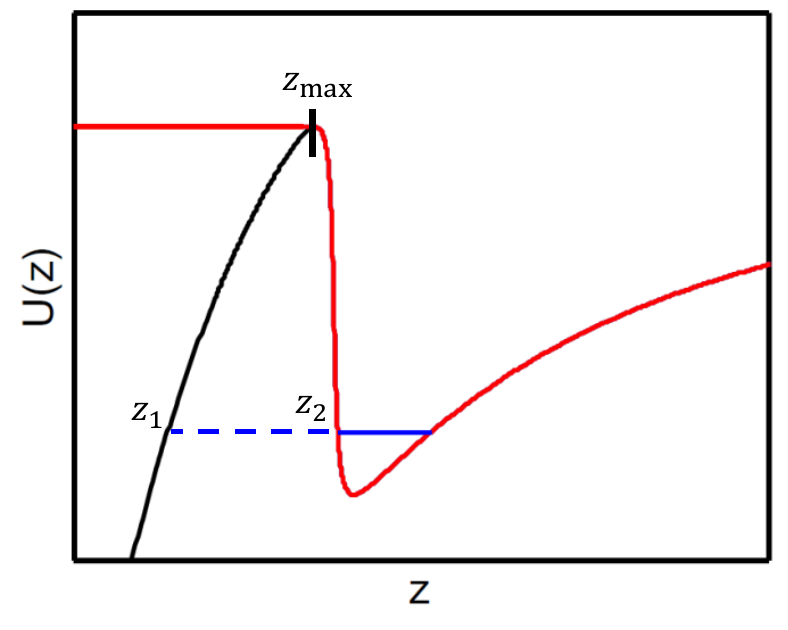}
\end{center}
\caption{Black curve: Metastable dressed state potential $U_d(z)$~(in arbitrary units) seen by the atom. The maximum of the metastable barrier is located at $z_{\footnotesize\textrm{max}}$. Red curve: Stable potential $U_s(z)$ used to approximately calculate the binding energy $\hbar\omega_b$~(blue line) of the atom in the trap. The classical turning points $z_{1,2}$ of a particle with this energy, which are used to evaluate the tunneling rate in the WKB approximation, are labeled as well.\label{fig:WKB}}
\end{figure}

The escape rate of an atom from the metastable potential formed by the vacuum trapping scheme due to quantum tunneling is calculated in the WKB approximation in two steps. First, we find an approximation to the real part of the bound state energy $\hbar\omega_b$. This is implemented by replacing the actual metastable dressed-state potential $U_d(z)$ with a stable one, $U_s(z)$, according to the formula~(also see Fig.~\ref{fig:WKB})
\bea U_s(z) & = & U_d(z) \;\;\;\;\; (z>z_{\footnotesize\textrm{max}}), \\
& = & U_d(z_{\footnotesize\textrm{max}}) \;\;\;\;\; (z\leq z_{\footnotesize\textrm{max}}). \eea
Here, $z_{\footnotesize\textrm{max}}$ is the position corresponding to the maximum of the metastable barrier. The ground-state energy of $U_s(z)$ is solved numerically.

Once the approximate ground-state binding energy is obtained, the tunneling probability $P_t$ for a particle hitting the metastable barrier is calculated using the WKB approximation,
\be P_t=4\exp\left(-2\int_{z_1}^{z_2} dz \sqrt{\frac{2m(U_d(z)-\hbar\omega_b)}{\hbar^2}}\right). \ee
Here $z_{1,2}$ are the solutions to $U_d(z)=\hbar\omega_b$ and are thus the classical turning points of a particle with energy $\hbar\omega_b$. The rate that the trapped atom collides with the barrier is approximated by $\Delta p_z/m \Delta z$, where the position and momentum uncertainties are obtained from the numerical ground-state solution. Thus we estimate the inverse of the tunneling-limited lifetime to be $\tau_{\footnotesize\textrm{tunnel}}^{-1}=\frac{P_t\Delta p_z}{m \Delta z}$.

\end{document}